\documentclass[onecolumn,showpacs,preprintnumbers,amsmath,amssymb]{revtex4}

\usepackage{graphicx}
\usepackage{dcolumn}
\usepackage{bm}

\begin{document}

\preprint{APS/123-QED}

\title{Does A Special Relativistic Liouville Equation Exist?}

\author{Jose A. Magpantay}
\email{jose.magpantay11@gmail.com}
\affiliation{Quezon City, Philippines}%
\date{\today}

\begin{abstract}
The Liouville equation, the starting point of non-relativistic, non-equilibrium classical statistical mechanics, is problematic in special relativity because of two problems. A relativistic Hamiltonian is claimed not to exist for interacting particles and the problem of what time to use since the particles will all have their own time as part of the space-time coordinate. In this paper, I look at this problem and surprisingly find that there is no special relativistic Liouville equation in 8N phase space, where N is the number of particles, because the canonical Hamiltonian is zero for both non-interacting and interacting particles. This is due to the parametrization symmetry in defining a single time for the Liouville equation evolution, which results in a constraint. This is similar to the fact that in general relativity, diffeomorphism invariance of the theory always give a zero Hamiltonian because of a constraint.
\end{abstract}

\pacs{Valid PACS appear here}
\maketitle


1. Non-relativistic, classical statistical mechanics starts with the Liouville equation for the Gibbs distribution. Consider a system of N particles with the Hamiltonian
\begin{equation}\label{1}
H_{nr} = \sum_{a=1}^{N} \dfrac{\vec{p}_{a}^{2}}{2m} + \sum_{a,b, b>a}^{N} V(\vert \vec{x}_{a} - \vec{x}_{b} \vert),
\end{equation}
where the two-body potential is instantaneous. The N particle Gibbs distribution $ f(\vec{x}_{a}, \vec{p}_{a}; t) $ is solved from the Liouville equation
\begin{equation}\label{2}
\dfrac{\partial f}{\partial t} - \left\lbrace  H_{nr}, f \right\rbrace  = 0.
\end{equation}
In non-relativistic physics, where interactions are instantaneous, the single time t in the Liouville equation is consistent. However, the above equation is hardly solvable for realistic potentials and contains too much information anyway so formally one follows the BBGKY hierarchy of equations till the process ends with the Boltzmann equation for the one particle distribution. This in turn is used to prove the H Theorem, which established the Second Law of Thermodynamics.

2. When the system is relativistic, two problems must be settled before a Liouville equation can be defined. The first is what is the Hamiltonian? The old theorem of (1) Currie, Jordan and Sudarshan \cite{Currie}; (2) Cannon and Jordan \cite{Cannon} and (3) Leutwyler \cite{Leutwyler} states there is no relativistic Hamiltonian for interacting particles because no Hamiltonian, the time translation generator, will satisfy the Poincare algebra. Without a Hamiltonian, there is no Liouville equation. 

In a previous paper \cite{Magpantay}, I proposed a relativistic extension of the non-relativistic Hamiltonian given by equation (1). The kinetic term is easily extended to the relativistic case and is given by $ \sqrt{p^{2}c^{2} + m^{2}c^{4}} $. The potential term is known for specific two-body interactions. For the Coulomb potential, the relativistic potential is the time-delayed potential given by
\begin{subequations}\label{3}
\begin{gather}
V_{r} \sim [\dfrac{1}{R(1 - \vec{v'}\cdot\vec{n})}]_{ret},\label{first}\\
R = \vert \vec{x}(t) - \vec{x'}(t') \vert, \label{second}\\
\vec{v'} = \dot{\vec{x'}}(t'), \label{third}\\
\vec{n} = \dfrac{\vec{x}(t) - \vec{x'}(t')}{R},
\end{gather} 
\end{subequations}
and ret means the potential acting on the charge at $ \vec{x} $ at t is due to the charge at $ \vec{x'} $ at earlier time t', which is solved from
\begin{equation}\label{4}
t - t' - \frac{R}{c} = 0.
\end{equation}

But what about other potentials? Is there a prescription on how to make relativistic the non-relativistic two-body potential given in Equation (1)? My previous work that I just referred to do just that in two steps as given next.
\begin{subequations}\label{5}
\begin{gather}
n^{2}(\vert \vec{k} \vert^{2}) = \left[ \int d^{3}x V(\vert \vec{x} \vert) \exp {i \vec{k} \cdot \vec{x} } \right]^{-1}, \label{first}\\
G_{4}(\vec{x}(t) - \vec{x'}(t'); t - t') = \int d^{3}k dk_{0} \dfrac{\exp {[ik_{0}(t - t') - i\vec{k} \cdot (\vec{x} - \vec{x'})]} }{n^{2}( \vert \vec{k} \vert^{2} - \frac{k_{0}^{2}}{c^{2}} )}, \label{second}\\
V_{r}(\vert \vec{x} - \vec{x'} \vert; t) = \frac{1}{2} \int dt' G_{4}(\vec{x}(t) - \vec{x'}(t'); t - t') 
\end{gather}
\end{subequations}
Equation (5a) gives the $ n^{2}(\vert \vec{k} \vert^{2}) $ from the Fourier transform of the two-body potential. Then this is extended to the argument from $ \vert \vec{k} \vert^{2} $ to  $ \vert \vec{k} \vert^{2} - \frac{k_{0}^{2}}{c^{2}} $ before substituting in equations (5b) to (5c) to give the relativistic two-body potential $ V_{r} $.  If we apply this prescription to the Coulomb potential, it will result in the time-delayed Coulomb potential given by equation (3).

For the case of N particles, where N is typically of the order of Avogadro's number ($ 10^{23} $), what time to use becomes a problem. The obvious solution is to have an observer O in the relativistic frame S that will specify (in principle) the position of all the particles at any particular time t, i.e., give $ \vec{x_{a}}(t) $ for $ a = 1,..N $. The observer is ascribed the capacity for isochronous specification of the positions of all the particles. But this is problematic for the simple reason that in another Lorentz frame, the isochronicity is lost, i.e., $ ct'_{a} = \gamma ( ct + \beta x_{a} ) $, and $\gamma = \dfrac{1}{\sqrt{1 - \beta^{2}}} $ and $ \beta = \frac{u}{c} $. Furthermore, the particles are moving with relativistic velocities, thus time delay in the times of $ N \propto 10^{23} $ particles will make it impossible for an observer at rest in S to specify isochronously the positions of all the particles.   

This is the first problem that must be resolved, what single time t do we use for the N particle system to go with the relativistic Hamiltonian. This single time t is important for the Hamiltonian dynamics is based on equal time Poisson brackets and the same for the Liouville equation. 

3. Before the choice of single time t is settled, a sense of what the relativistic Hamiltonian that corresponds to the non-relativistic Hamiltonian given by equation (1) must be explored. In a previous paper, I proposed the Hamiltonian
\begin{equation}\label{6}
H_{r} = \sum_{a} (\vec{p}_{a}^{2}c^{2} + m^{2}c^{4})^{\frac{1}{2}} + \frac{1}{2} \sum_{a,b; b>a} \int dt_{b}' G_{4}(\vert \vec{x}_{a} - \vec{x}_{b}\vert; t_{a} - t_{b}).
\end{equation}
Although $ H_{r} $ given by equation (6) does not satisfy the Poincare algebra because of the time-delay, it seems to be a legitimate relativistic Hamiltonian by construction - the kinetic term is the relativistic generalization of the non-relativistic kinetic energy and the potential energy term is the relativistic generalization of the instantaneous two-body potential. And as already discussed after equation (5), the $ G_{4} $ term gives the retarded potential for the Coulomb problem. 

Thus, it seems that the relativistic Hamiltonian given by equation (6) is a Hamiltonian for the relativistic Liouville equation version of equation (2). The problem is equation (6) involves N times $ t_{a} $ in the momentum terms as shown by $ \vec{p}_{a} = \dfrac{m\vec{v}_{a}}{\sqrt{1 - \frac{\vert\vec{v}_{a}\vert^{2}}{c^{2}}}} $, where $ \vec{v}_{a} = \frac{d\vec{x}_{a}}{dt_{a}} $ and more explicitly in the $ G_{4} $ term. The problem then is to express equation (6) in terms of a single time t required of Hamiltonian dynamics. And imploring an observer in S capable of an isochronous time for all N particles is not relativistically consistent as argued in Section 2. 

3. The obvious approach to the single time problem is to begin with the action and the Lagrangian, introduce a single time parameter and then proceed via the canonical procedure 
\begin{subequations}\label{7}
\begin{gather}
S_{r} = \sum_{a} \int dt_{a} L_{r}(a), \label{first}\\
L_{r}(a) = (-mc^{2})\left( 1 - \frac{1}{c^{2}} \dfrac{d\vec{x}_{a}}{dt_{a}} \cdot \dfrac{d\vec{x}_{a}}{dt_{a}} \right)^{\frac{1}{2}} - \frac{1}{2} \sum_{b \neq a} \int dt_{b} G_{4}\left(  \vec{x}_{a} - \vec{x}_{b}; t_{a} - t_{b} \right).
\end{gather}
\end{subequations}
The first term of $ L_{r}(a) $ is the free relativistic particles term and the second the relativistic two-body interaction term. 

Treating as dynamical degrees of freedom all the space-time coordinates $ x_{a \mu} = ( ct_{a},\vec{x}_{a} ) $, which evolve according to the mathematical (not measured by a clock) time parameter t, common to all particles, the action and Lagrangian given by equation (7) becomes
\begin{subequations}\label{8}
\begin{gather}
S_{r} = \int dt L_{r}, \label{first}\\
\begin{split}
L_{r}& = \sum_{a} \Big\lbrace (-mc) \left( \eta_{\mu \nu} \dfrac{dx_{a \mu}}{dt}  \dfrac{dx_{a \nu}}{dt} \right)^{\frac{1}{2}} \\
& \quad - \frac{1}{2c^{2}} \dfrac{dx_{a 0}}{dt} \sum_{b \neq a} \int dt' \dfrac{dx_{b 0}}{dt'} G_{4}( \vec{x}_{a} - \vec{x}_{b}; \frac{1}{c} ( x_{a 0}(t) - x_{b 0}(t') ) )\Big\rbrace.
\end{split}
\end{gather}
\end{subequations}
Since the action and Lagrangian is in terms of a single time t, except for the time-delayed two-body interaction term given by a single time delay t', then the canonical procedure will give the appropriate Hamiltonian, which can be used in the Liouville equation. 

For simplicity, take first the non-interacting case, i.e., put $ G_{4} = 0 $. The conjugate momenta are
\begin{subequations}\label{9}
\begin{gather}
\vec{p}_{a} = mc \dfrac{d\vec{x}_{a}}{dt} \left[ (\dfrac{d{x}_{a 0}}{dt})^{2} - (\dfrac{d\vec{x}_{a}}{dt})^{2} \right]^{-\frac{1}{2}}, \label{first}\\
\pi_{a} = -mc\frac{dx_{a 0}}{dt} \left[ (\dfrac{d{x}_{a 0}}{dt})^{2} - (\dfrac{d\vec{x}_{a}}{dt})^{2} \right]^{-\frac{1}{2}}.
\end{gather}
\end{subequations}
Unfortunately, equations (9; a,b) will not solve for the velocities in terms of the momenta and coordinate. However, equation (9 a) gives
\begin{equation}\label{10}
\dfrac{d\vec{x}_{a}}{dt} = \pm \dfrac{\vec{p}_{a}} {( m^{2}c^{2} + p_{a}^{2} )^{\frac{1}{2}}} \dfrac{dx_{a 0}}{dt},
\end{equation}
essentially relating the two velocities. Substituting this in equation (9 b), a constraint given by
\begin{equation}\label{11}
\pi_{a} \pm (p_{a}^{2} + m^{2}c^{2})^{\frac{1}{2}} = 0,
\end{equation} 
arises.

Solving for the Hamiltonian gives
\begin{equation}\label{12}
\begin{split}
H_{r}& = \sum_{a} \left( \vec{p}_{a} \cdot \frac{d\vec{x}_{a}}{dt} + \pi_{a} \frac{dx_{a 0}}{dt} \right) - L_{r}\\
       & = \sum_{a} \left[ \pi_{a} \pm \left( p_{a}^{2} + m^{2} c^{2} \right)^{\frac{1}{2}} \right] \frac{dx_{a 0}}{dt},
\end{split} 
\end{equation}
which makes the Hamiltonian equal to zero by virtue of the constraint given by equation (11).

This reminds of the situation in gravity, where the Hamiltonian is zero because of a constraint, which is a consequence of diffeomorphism symmetry. What is the equivalent in this case? Note that the free relativistic particles action as given by equation (8) (putting $ G_{4} = 0 $) is Lorentz invariant provided the time t is a Lorentz scalar. However, the action is invariant under reparametrization, i.e., $ t \rightarrow t' = t'(t) $. This symmetry yields the constraint given by equation (11) resulting in the zero Hamiltonian.

Note, the reparametrization symmetry only starts for 2 or more particles. With a single particle, the Lagrangian is just the $ a = 1 $ of equation (7), thus there is no need to introduce a new time t since the relativistic time of the particle naturally describes the time evolution of the particle and the Hamiltonian is simply $ \sqrt{( p^{2} c^{2} + m^{2} c^{4} )} $. This result is used to say that the Hamiltonian for N non-interacting relativistic particles is the first term of the Hamiltonian given by equation (6). But as already pointed out in Section 3, this is erroneous for it involves N relativistic times $ t_{a}, a = 1,..., N $. And in Hamiltonian dynamics, Poisson brackets involve equal times with only a single time of evolution. Thus, the result of this paper, that the single time Hamiltonian even for non-interacting relativistic particles vanish is rather surprising.

4. Now, the time-delayed two-body interaction is also considered (the $ G_{4} $ term). From the full Lagrangian given by equation (9), the conjugate momenta are
\begin{subequations}\label{13}
\begin{gather}
\vec{P}_{a} = mc \dfrac{d\vec{x}_{a}}{dt} \left[ (\dfrac{dx_{a 0}}{dt})^{2} - (\dfrac{d\vec{x}_{a}}{dt})^{2} \right]^{-\frac{1}{2}}, \label{first}\\
\Pi_{a} = -mc\dfrac{dx_{a 0}}{dt} \left[ (\dfrac{dx_{a 0}}{dt})^{2} - (\dfrac{d\vec{x}_{a}}{dt})^{2} \right]^{-\frac{1}{2}} - \frac{1}{2c^{2}} \sum_{b \neq a} \int dt' G_{4}\left( \vec{x}_{a} - \vec{x}_{b}; \frac{1}{c} ( x_{a 0}(t) - x_{b 0}(t') ) \right) \dfrac{dx_{b 0}}{dt'}.
\end{gather}
\end{subequations}
There is a subtlety in how equation (13 b) is arrived at. There is an extra term given by
\begin{equation}\label{14}
\sum_{b,c; b \neq c} \dfrac{dx_{b 0}}{dt} \int dt' G_{4}(\vec{x}_{b} - \vec{x}_{c}; \frac{1}{c} \left[ x_{b 0}(t) - x_{c 0}(t') \right] ) \dfrac{\delta \frac{dx_{c 0}(t')}{dt'}}{\delta\frac{dx_{a 0}(t)}{dt}}.
\end{equation}
The last term in this equation is $ \delta(t - t') $ and by virtue of $ G_{4} $ encoding time-delay in $ t_{a}(t) - t_{b}(t') $, then t will never equal t' making the term given by equation (14) zero.

Equation (13 a) gives exactly the same equation as Equation (10), only $ \vec{p}_{a} $ is replaced by $ \vec{P}_{a} $. Substituting this equation in Equation (13 b), gives
\begin{equation}\label{15}
\Pi_{a} \pm (P_{a}^{2} + m^{2}c^{2})^{\frac{1}{2}} + \frac{1}{2c^{2}} \sum_{b \neq a} \int dt' G_{4}\left( \vec{x}_{a} - \vec{x}_{b}; \frac{1}{c} ( x_{a 0}(t) - x_{b 0}(t') ) \right) \dfrac{dx_{b 0}}{dt'} = 0.
\end{equation}
Equation (15) seems not to be constraint, it appears to solve for the velocity $ \dfrac{dx_{a 0}}{dt} $, in terms of the momenta and coordinates, which will then lead to an expression for the velocity $ \dfrac{d\vec{x}_{a}}{dt} $. Unfortunately, the integral equation is not solvable. However, equation (15) is a constraint. The integral involves a velocity, $ \dfrac{dx_{b 0}}{dt'} $ with coordinate coefficients in $ G_{4} $, thus it will lead to a function of coordinates. Another way to see this is that equation (15) is really a coordinate integral, i.e., $  \int G_{4}\left( \vec{x}_{a} - \vec{x}_{b}; \frac{1}{c} ( x_{a 0}(t) - x_{b 0}(t') ) \right) dx_{b 0} $, thus resulting in a function of coordinates only. This will make equation (15) a constraint.

The Hamiltonian can be computed giving
\begin{equation}\label{16}
\textbf{H}_{r} = \sum_{a} \left\lbrace   \Pi_{a} \pm (P_{a}^{2} + m^{2}c^{2})^{\frac{1}{2}} + \frac{1}{2c^{2}} \sum_{b \neq a} \int dt' G_{4}\left( \vec{x}_{a} - \vec{x}_{b}; \frac{1}{c} ( x_{a 0}(t) - x_{b 0}(t') ) \right) \frac{dx_{b 0}}{dt'} \right\rbrace \dfrac{dx_{a 0}}{dt},
\end{equation}
and by equation (15), the Hamiltonian is exactly zero. 

The vanishing of this Hamiltonian again follows from reparametrization invariance of the action given by equation (8; a,b). The reparametrization invariance of the free particles part have been shown in the previous section. The reparametrization invariance of the two-body interaction term follows similarly.

5. In this paper, I showed that in 8N phase space, the relativistic Hamiltonian for a system with two-body interaction given by the time-delayed $ G_{4} $ is zero. This means a relativistic Liouville equation does not exist. 

The result is consistent with the old theorems in the 1960s that claim that there is no relativistic Hamiltonian for interacting relativistic particles. In this paper, I showed it is zero by introducing a mathematical one time parameter t in the relativistic Lagrangian given by equation (7; a,b) resulting in a one time Lagrangian given by equation (8; a,b), which is now used in the canonical procedure.

The surprising result is that even for free relativistic particles, which generally assumes the Hamiltonian given by the first term of equation (6), the 8N Hamiltonian with a single time t is zero. The reason that the Hamiltonian given by the first term of equation (6) is not valid for Hamiltonian dynamics and Liouville equation is that it still involves N times $ t_{a}, a = 1,...N $. And introducing a single time t, which results in equation (7) turning into a single time action given by equation (8), and following the canonical procedure results in the 8N phase space Hamiltonian vanishing because of the reparametrization symmetry. 

The result arrived at in this paper is a bit difficult to accept because it means that the Liouville approach to relativistic many particles does not exist. There must be a way out of this problem. The transition from equation (7) to (8) where the mathematical single time t was introduced, suggests as the source of this null result. The solution to this problem is to introduce one of the physical times say $ t_{1} $, as the Liouville time and the times of the other particles solved from the time delay coming from $ G_{4}(\vec{x}_{1} - \vec{x}_{b}; t_{1} - t_{b} ) $, with $ b \neq 1 $. But this is rather difficult to implement.     

\begin{acknowledgments}
I would like to thank Mike Solis of the National Institute of Physics of UP Diliman for reading the manuscript and raising questions. To David, my first apo, I dedicate this paper. His pictures brought me joy during this pandemic.
\end{acknowledgments}

\end{document}